\documentclass{article}

\usepackage{graphicx} 
\usepackage[utf8]{inputenc}
\usepackage[T1]{fontenc}
\usepackage[hidelinks]{hyperref}
\usepackage{url}
\usepackage{times}
\usepackage{lmodern}
\usepackage{xcolor}
\usepackage{booktabs} 
\usepackage{amssymb}
\usepackage{amsmath}

\usepackage{multirow}%
\usepackage{amsmath,amssymb,amsfonts}%
\usepackage{amsthm}%
\usepackage{mathrsfs}%
\usepackage[title]{appendix}%
\usepackage{xcolor}%
\usepackage{textcomp}%
\usepackage{manyfoot}%
\usepackage{booktabs}%
\usepackage{algorithm}%
\usepackage{algorithmicx}%
\usepackage{algpseudocode}%
\usepackage{listings}%
\usepackage{bm}%
\usepackage{siunitx}%
\usepackage{orcidlink}
\usepackage[document]{ragged2e}

\DeclareMathOperator{\DA}{DA}  
\DeclareMathOperator{\sgn}{sgn}  

\usepackage[style=apa,
citestyle=authoryear,
sorting=nyt,
sortcites=true,
autopunct=true,
babel=hyphen,
hyperref=true,
abbreviate=false,
backref=false,
dashed=false,
maxcitenames=2, 
apamaxprtauth=999, 
uniquename=false,
uniquelist=false,
backend=biber]{biblatex}

\title{\vspace{-2cm} Performance of models for monitoring sustainable development goals from remote sensing: A three-level meta-regression}

\author{
	Jonas Klingwort$^{1,\ast,\orcidlink{0000-0002-4545-9136}}$, Nina M. Leach$^{2,\orcidlink{0000-0002-6398-7522}}$, Joep Burger$^{1,\orcidlink{0000-0002-7298-5561}}$\\ 
	\\
	\normalsize{$^{1}$Department of Research \& Development, Statistics Netherlands (CBS)}\\ 
    \normalsize{$^2$ Department of Methodology and Statistics, Leiden University}\\
	\normalsize{$^\ast$correspondence: j.klingwort@cbs.nl}
    \vspace{-1cm}
}
\date{~}

\begin{document}
	
	\maketitle
	
	\begin{abstract}
		\noindent \textbf{Abstract:}
		Machine learning (ML) is a tool to exploit remote sensing data for the monitoring and implementation of the United Nations' Sustainable Development Goals (SDGs). In this paper, we report on a meta-analysis to evaluate the performance of ML applied to remote sensing data to monitor SDGs. Specifically, we aim to 1) estimate the average performance; 2) determine the degree of heterogeneity between and within studies; and 3) assess how study features influence model performance. Using the PRISMA guidelines, a search was performed across multiple academic databases to identify potentially relevant studies. A random sample of 200 was screened by three reviewers, resulting in 86 trials within 20 studies with 14 study features. Overall accuracy was the most reported performance metric. It was analyzed using double arcsine transformation and a three-level random effects model. The average overall accuracy of the best model was 0.90 $[0.86, 0.92]$. There was considerable heterogeneity in model performance, 64\% (third-level $I^2$) of which was between studies. The only significant feature was the prevalence of the majority class, which explained 61\% (third-level $R^2$) of the between-study heterogeneity. None of the other thirteen features added value to the model. The most important contributions of this paper are the following two insights. 1) Overall accuracy is the most popular performance metric, yet arguably the least insightful. Its sensitivity to class imbalance makes it necessary to normalize it, which is far from common practice. 2) The field needs to standardize the reporting of model performance. Reporting of the confusion matrix for independent test sets is the most important ingredient for between-study comparisons of ML classifiers. These findings underscore the critical need for robust and comparable evaluation metrics in machine learning applications to ensure reliable and actionable insights for effective SDG monitoring and policy formulation.
	\end{abstract}
	
	
	\section{Introduction}\label{sec:intro}

In 2015, all United Nations member states adopted the Sustainable Development Goals (SDGs) to address global challenges such as climate change, environmental degradation, poverty, and inequality \parencite{undesa2023,unggim2019,un2024}. This international plan outlines 17 global goals for a better and more sustainable future. Having passed the midpoint of the SDGs' timeline with significant setbacks, the critical role of timely and high-quality data has never been more apparent. These data are vital to identifying challenges, formulating evidence-based solutions, monitoring the implementation of solutions, and making essential course corrections. Traditional monitoring approaches, such as household- or field-level surveys (ground-acquired data), remain the primary source of data collection for key indicators of SDGs by National Statistical Institutes (NSIs) \parencite{burke2021}. These methods are expensive and time-consuming to conduct. As a result, the frequency of ground-acquired data varies significantly worldwide; for example, the most recent agricultural census for 24\% of the world's countries was more than 15 years ago \parencite{burke2021}. Recognizing this challenge, both \textcite{undesa2023} and \textcite{sigd2017} underscore the importance of innovative methodology and data sources, including remote sensing and ML, to enhance the monitoring and implementation of the SDGs.

Remote sensing data are collected from a distance via satellite, aircraft, or drones. They offer a cost-effective approach for monitoring wide-ranging geographic areas \parencite{khatami2016,maso2023,unggim2019,zhao2022}. Remote sensing imagery has been limited to agricultural and socioeconomic applications for decades \parencite{burke2021,lavallin2021,zhang2022a}. For instance, the Laboratory for Applications of Remote Sensing (LARS) has utilized satellite data and ML methods for crop identification since the 1960s \parencite{holloway2018}. However, in recent years, there has been a considerable increase in the spatial, spectral, and temporal resolution of remote sensing data, alongside a significant increase in free sensor data and computational power for complex data analysis \parencite{burke2021,thapa2023,zhang2022a}. The magnitude of possible applications and increased availability of remote sensing data have rapidly increased the number of published research papers in this field \parencite{burke2021,khatami2016}. Earth observation satellites alone can measure 42\% of the SDG targets \parencite{zhang2022a}. Copernicus, the earth observation program of the European Union, provides services on land monitoring, emergency management, atmosphere monitoring, marine environment monitoring, climate change, and security.

Earth observation data has multiple processing levels. Raw data contain spectral radiance (mW mm\textsuperscript{-2} sr\textsuperscript{-1} nm\textsuperscript{-1}) that needs correction and calibration. Earth systems models are applied to derive physical, chemical, and biological parameters. More modeling is involved to combine data over space and time. ML can be applied to learn relationships between processed earth observation data and labels derived from linking register data or manual annotation. This paper is about the performance of these ML models for monitoring SDGs.

Given the wide variety of methodologies and contexts in previous studies, a critical question arises: what factors influence the performance of ML models using remote sensing data for SDG monitoring? A meta-analysis statistically combines the body of evidence on a specific topic to produce unbiased summaries of evidence \parencite{iliescu2022}.

This study seeks to address the question of how ML models perform when applied to remote sensing data for SDG monitoring. By conducting a meta-analysis on peer-reviewed research articles in this domain, the study aims to 1) estimate the average performance (population effect size), 2) determine the degree of heterogeneity between and within studies, and 3) assess whether specific study features can explain model performance and heterogeneity using a meta-regression. This approach enables consistent, transparent, and reproducible assessments across studies, which is crucial for informing decision-makers and stakeholders involved in SDG monitoring. By accounting for methodological heterogeneity and contextual factors, our approach supports prioritizing ML models that are suited for deployment in real-world and policy-relevant applications.

Several studies have previously aggregated research on the application of remote sensing for SDG monitoring. These reviews either apply methodology that aligns more closely with synthesis without meta-analysis \parencite{campbell2020}---for example, \textcite{thapa2023} and \textcite{ekmen2024}---or apply unweighted meta-analysis techniques, such as \textcite{khatami2016} and \textcite{hall2023}. In an unweighted meta-analysis, all studies are treated equally regardless of their sample size, quality, or variance \parencite{hall2018}. Our study uses a random effects approach by including the sample sizes of the primary studies when aggregating the results of the individual studies. This study addresses a specific research gap: a weighted multi-level meta-analysis of a proportion in the domain of machine learning for monitoring SDGs from remote sensing. We refer to, for example, \textcite{holloway2018,Estoque2020May} for general literature reviews on ML and remote sensing.

The key findings observed are that we had to focus on overall accuracy as this is the most commonly reported performance metric. The only significant feature was the prevalence of the most common class, which explained most of the between-study heterogeneity. This finding shows the limited utility of overall accuracy and the need for reporting of more standardized quality metrics. The originality and new contributions of this paper in the field of remote sensing and sustainable development goals is the combination of a meta-analysis on a proportion taking into account the hierarchy of the data.

The paper is organized as follows: Section \ref{sec:background} provides the research background on remote sensing and ML. The methods used are explained in Section \ref{sec:methods}. The results are presented in Section \ref{sec:results}. The study is discussed and concluded in Section \ref{sec:discussion}.

\section{Research background}\label{sec:background}

This section provides an overview of the concepts and methodologies. It introduces remote sensing and ML techniques.

\subsection{Remote sensing}

\begin{figure}[htbp]
    \centering
    \includegraphics[width=0.5\linewidth]{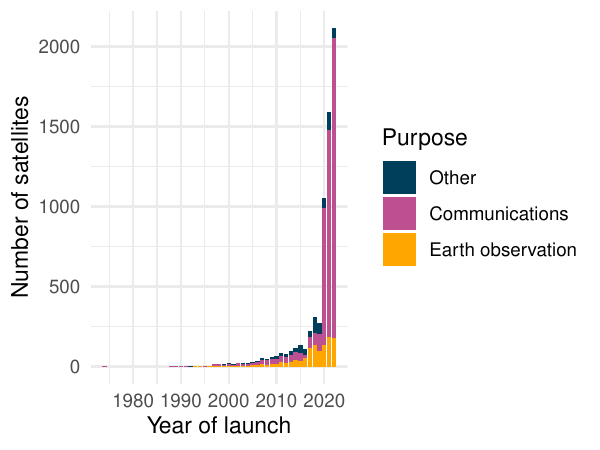}
    \caption{Number of satellites launched per year by purpose. Data from \cite{ucs2023}.}
    \label{fig:nsat}
\end{figure}

In the broadest sense, remote sensing involves acquiring information about an object or phenomenon without direct contact \parencite{campbell2011}. More specifically, remote sensing refers to gathering data about land or water surfaces using sensors mounted on aerial or satellite platforms that record electromagnetic radiation reflected or emitted from the earth's surface \parencite{campbell2011}. The origins of remote sensing lie with the development of photography in the 19th century, with the earliest aerial or earth observation photographs taken with cameras mounted on balloons, kites, pigeons, and airplanes \parencite{burke2021,campbell2011}. The first mass use of remote sensing was during World War I with aerial photography. The modern era of satellite-based remote sensing started with the launch of Landsat 1 in 1972, the first satellite specifically designed for earth observation \parencite{campbell2011}. Today, remote sensing technology enables frequent and systematic data collection about the earth's surface with global coverage, revolutionizing our ability to monitor and analyze the earth's surface \parencite{burke2021,nasa2024}. 
The number of satellites that are launched every year increases over time, especially for communications purposes (like the Starlink satellites), but also for earth observation (Fig. \ref{fig:nsat}).

Sensors on remote sensing devices such as satellites measure electromagnetic radiation reflected by objects on the earth's surface. This is done in two different ways: passive and active. Passive sensors rely on natural sources like sunlight to record incident energy reflected off the earth's surface. Active sensors generate their energy, which is emitted and then measured as it reflects from the earth's surface \parencite{nasa2024}.

The earth's surface components have different spectral signatures---i.e., reflect, absorb, or transmit energy in different amounts and wavelengths \parencite{campbell2011}. Remote sensing devices have several sensors that measure specific ranges of wavelengths in the electromagnetic spectrum; these are referred to as spectral bands, e.g., infrared, visible light, or ultraviolet \parencite{nasa2024}. By capturing information from particular bands, the spectral signatures of surfaces can be used to identify objects on the ground.

\subsection{Machine learning}
\label{sec:ML}

ML is a research field focused on developing and applying algorithms that can learn to find patterns in data. ML techniques can be applied to different types of data. If the data are unlabeled, i.e., they contain only features but no variable of interest, unsupervised ML techniques can be applied to group instances (clustering) or features (dimension reduction). If the data are labeled, i.e., they contain both features and a variable of interest, supervised ML techniques can be applied to map the relations between features and labels and to predict the variable of interest for unlabeled instances. If the variable of interest is categorical, ML solves a classification problem. If the variable of interest is numerical (discrete or continuous), ML solves a regression problem. In this paper, the focus is on ML classifiers.

ML classifiers such as neural networks (NN), random forests (RF), and support vector machines (SVM) have long been applied for spatial data analysis and geographic modeling \parencite{lavallin2021}. For NNs we refer to \textcite{goodfellow2016,Zhu2017Dec,ma2019,Reichstein2019Feb}, for RFs we refer to \textcite{breiman2001,Rodriguez-Galiano2012Jan,Belgiu2016Apr,Strobl2007Dec,cutler2017}, and for SVMs to \textcite{vapnik1995,Huang2002Jan,Melgani2004Aug,Mountrakis2011May}. We refer to \textcite{Kavvada2020Sep} who give a broad overview of applications to each SDG goal in which these methods are used. For readers interested in the mathematical representation of these models, we refer to the standard textbooks by \textcite{james2021,bishop2006,hastie2009}. For recent reviews on Artificial Intelligence techniques used in the context of the SDGs, we refer to \cite{su15,su151813493}.

Compared to using spectral indices (e.g., vegetation index) alone, ML techniques enhance the accuracy and efficiency of data analysis and interpretation processes, making it possible to analyze large volumes of data effectively. This is particularly useful for handling remote sensing data's high complexity and dimensionality. In recent years, the application of ML techniques in remote sensing has surged, driven by the increasing availability of large datasets and advancements in computational power \parencite{unggim2019,zhang2022a}.

Performance metrics are used to test the quality of model predictions on independent test sets. For classification tasks, these are derived from a confusion matrix: a cross-tabulation of the number of instances observed in class $r \in \{ 1, \cdots, q \}$ and predicted in class $c \in \{ 1, \cdots, q \}$, where $q$ is the number of classes (Table \ref{tab:confmatrix}). In a confusion matrix, the correctly classified instances are on the diagonal, and the off-diagonal cells indicate which classes are confused, i.e., incorrectly classified. In remote sensing applications, an instance can be a pixel, a group of pixels, or an image \parencite{stehman2019}.

\begin{table}[htbp]
\centering
  \caption{Confusion matrix for a single estimate of overall accuracy based on the classification of $n$ instances.}
  \label{tab:confmatrix}
  \begin{tabular}{l r l l l l l}
    \cmidrule{1-6}
    & & \multicolumn{3}{l}{Predicted} & & \\
    & Class & $1$ & $\cdots$ & $q$ & Sum & \\
    \cmidrule{1-6}
    Observed & $1$ & $n_{11}$ & $\cdots$ & $n_{1q}$ & $n_{1.}$ & $R_1$\\
    & $\vdots$ & $\vdots$ & $\ddots$ & $\vdots$ & $\vdots$ & $\vdots$ \\
    & $q$ & $n_{q1}$ & $\cdots$ & $n_{qq}$ & $n_{q.}$ & $R_q$\\
    & Sum & $n_{.1}$ & $\cdots$ & $n_{.q}$ & $n$ & \\
    \cmidrule{1-6}
    & & $V_1$ & $\cdots$ & $V_q$ & & $p$
  \end{tabular}
\end{table}

\noindent From this matrix, several performance measures can be derived \parencite{fao2016,stehman2019,unggim2019}. The simplest is overall accuracy, which is the fraction of instances that are correctly classified ($p = k / n$, where $k = \sum_{r=1}^{q} n_{rr}$). This metric is not class-specific. Two class-specific metrics are recall (aka sensitivity or producer's accuracy) and precision (aka reliability or user's accuracy). Recall of class $r$ is the fraction of instances observed in class $r$ that are correctly predicted ($R_r = n_{rr} / n_{r.}$). The precision of class $c$ is the fraction of instances predicted in class $c$ that are observed as such ($V_c = n_{cc} / n_{.c}$). F1 and average precision are two class-specific metrics that balance the trade-off between recall and precision of a class. The arithmetic or geometric mean recall and the area under the ROC curve ($R_r$ against $R_{\neg r}$) are metrics that balance the trade-off between recalls. It is important to realize that many performance metrics can be derived from a confusion matrix, that they serve different purposes, and that most are sensitive to class imbalance (the deviation from a discrete uniform distribution).

\section{Methods}\label{sec:methods}

\subsection{Article selection}
\label{sec:selection}

The methods adopted in this study are described in steps following the framework proposed by \textcite{debray2017}. In addition, efforts were made to follow the Preferred Reporting Items for Systematic Reviews and Meta-Analyses (PRISMA) guidelines \parencite{page2021}, although strict adherence was not always possible. In the following, we describe the criteria for article selection in detail.

\begin{figure}[t!]
    \centering
    \includegraphics[width=0.8\linewidth]{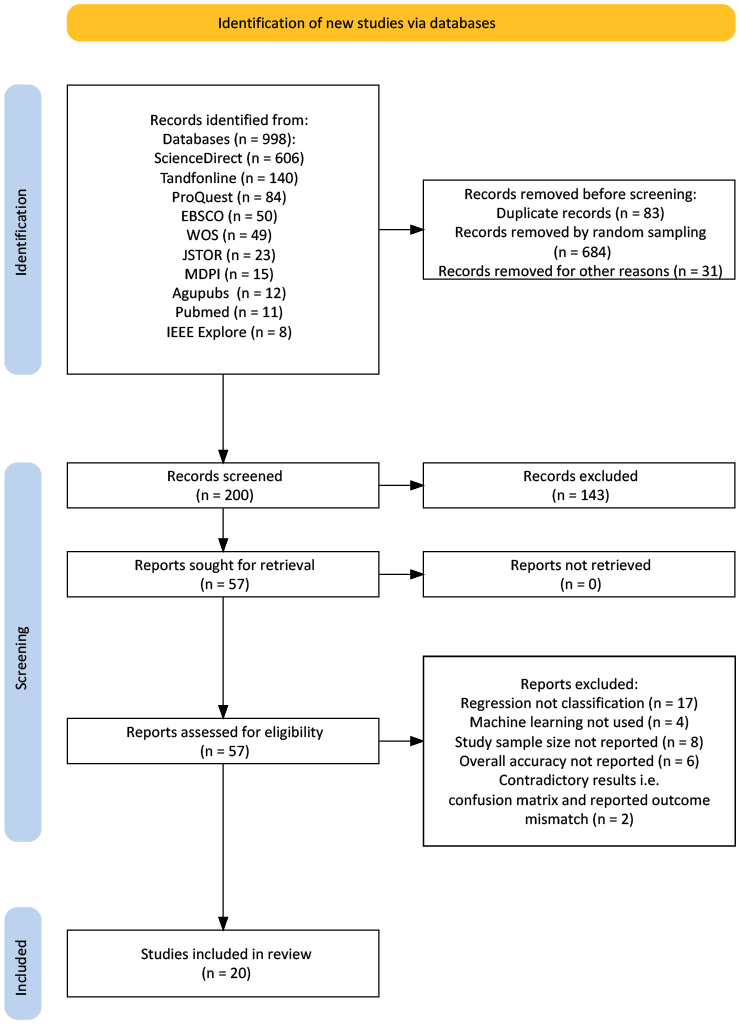}
    \caption{PRISMA flow diagram. Search and study selection process.}
    \label{fig:prisma}
\end{figure}

Peer-reviewed articles published between January 2018 and December 2023 were gathered on January 15 and 16, 2024, from several academic databases, including ScienceDirect and Taylor \& Francis Online (Fig. \ref{fig:prisma}). Several academic databases were used to reduce the potential bias from database coverage \parencite{hansen2022,tawfik2019}. Although Google Scholar can be useful for supplementary searches and gray literature, it is generally considered unsuitable as the primary source for systematic reviews \parencite{gusenbauer2020}. Furthermore, Google Scholar search results are not fully reproducible \parencite{gusenbauer2020}, and search result references cannot be downloaded in batches. Therefore, it was decided not to use Google Scholar to search for papers.

For the database query, three search terms were chosen for full-text searches: `remote sensing' AND `machine learning' AND `sustainable development goals'. These have been selected because it was assumed that papers addressing modeling approaches for monitoring sustainable development goals from remote sensing will likely include these terms. Initially, we experimented with including the common acronyms ‘ML’ and ‘SDGs’. However, it was ultimately decided to exclude these from the search terms because their inclusion increased the rate of irrelevant results (Considering ScienceDirect, including the acronyms increased the number of papers in 2018--2023 from 606 to 1064, representing a 76\% increase. For example, the abbreviation `ML' yields additional papers that deal with `Maximum Likelihood' estimation. The abbreviation `SDG' can also mean Standard Deviation of the Gradient (in some optimization literature) or Subsampled Double Gradient (in specific algorithms)). Consequently, only the three search terms mentioned at the beginning of this paragraph and no other search strings were used. This search resulted in 998 papers. The search results from these databases were downloaded in RIS format and imported into Zotero for further processing. Duplicate articles were handled using Zotero's `merge duplicates' function.

Out of the 998 papers, a simple random sample of 200 papers was drawn for title and abstract screening. All three authors independently screened these potentially relevant papers using the R package \texttt{metagear} \parencite{lajeunesse2016}. Keywords highlighted by the \texttt{metagear} user interface to speed up the screening process are given in Table \ref{tab:keywords}. The papers were selected according to the following criteria: a) publications utilizing remote sensing and ML techniques, and (b) indication of a quality assessment, such as overall accuracy. Of the 200 abstracts screened, only 57 were deemed potentially relevant by all three reviewers.

\begin{table}[htbp]
    \caption{Keywords highlighted by \texttt{metagear} during title and abstract screening.}
    \label{tab:keywords}
        \resizebox{\textwidth}{!}{

    \begin{tabular}{l p{0.8\linewidth}}
        \toprule
        Category & Keyword \\
        \midrule
        General & analysis, empirical, predictive, result, sustainable development \\
        Data & earth observation, remote sensing, remotely sensed, satellite \\
        Models & ann, bagging, bayes, boosting, cart, classification, classifier, decision tree, deep learning, gradient, ML, neural network, random forest, regression, regression tree, rf, supervised, support vector machine, svm, test set, training set \\
        Quality metrics & accuracy, auc, coefficient of determination, error, f1, mae, mean absolute error, mse, overall accuracy, precision, recall, rmse, roc, sensitivity, specificity \\
        To omit & meta-analysis, review, systematic review \\
        \bottomrule
    \end{tabular}}
\end{table}

To have comparable performance metrics, it was decided to focus on papers related to classification. The titles and abstracts of the 57 articles were screened using \texttt{metagear} to divide them into classification (40) and regression (17) papers. The papers using regression were excluded because they require different quality metrics. Two studies were excluded because the reported confusion matrices did not match the reported results and therefore deemed unreliable. Studies with incomplete reporting on study-level features were kept and the feature was coded `unknown' (see Section \ref{sec:explore}).

\noindent In the 40 classification papers, overall accuracy was the most commonly reported performance metric, and therefore, it was decided to include all papers that report overall accuracy. Only half of the selected classification papers reported the necessary information for the meta-regression. Most papers ($j = 1, \dotsc, h$) report several ($m_j$) estimates of overall accuracy from different trials, resulting in $m = 87$ estimates of overall accuracy from $h = 20$ papers (Table \ref{tab:papers}). $p_{ij}$ is the observed overall accuracy from trial $i$ in study $j$. The hierarchical structure of the data is addressed in the subsection on statistical analysis.

\begin{table}[t!]
    \caption{Selected studies sorted by number of trials per study and year of publication.}
    \label{tab:papers}
    \centering
    \resizebox{\textwidth}{!}{
    \begin{tabular}{rlrl}
        \toprule
        Study $j$ & Reference & Number of trials, $m_j$ & $p_{ij}$ \\
        \midrule
        1 & \textcite{nazir2020} & 27 & $p_{1,1}, \dotsc, p_{27,1}$ \\
        2 & \textcite{shen2024} & 7 & $p_{1,2}, \dotsc, p_{7,2}$ \\
        3 & \textcite{priyanka2023} & 7 & $p_{1,3}, \dotsc, p_{7,3}$ \\
        4 & \textcite{jochem2018} & 7 & $p_{1,4}, \dotsc, p_{7,4}$ \\
        5 & \textcite{solorzano2023} & 6 & $p_{1,5}, \dotsc, p_{6,5}$ \\
        6 & \textcite{aja2022} & 6 & $p_{1,6}, \dotsc, p_{6,6}$ \\
        7 & \textcite{agrillo2021} & 4 & $p_{1,7}, \dotsc, p_{4,7}$ \\
        8 & \textcite{shaharum2020} & 3 & $p_{1,8}, \dotsc, p_{3,8}$ \\
        9 & \textcite{tao2019} & 3 & $p_{1,9}, \dotsc, p_{3,9}$ \\
        10 & \textcite{aldousari2023} & 2 & $p_{1,10}, p_{2,10}$ \\
        11 & \textcite{fagua2023} & 2 & $p_{1,11}, p_{2,11}$ \\
        12 & \textcite{li2022} & 2 & $p_{1,12}, p_{2,12}$ \\
        13 & \textcite{xia2022} & 2 & $p_{1,13}, p_{2,13}$ \\
        14 & \textcite{mashaba2021} & 2 & $p_{1,14}, p_{2,14}$ \\
        15 & \textcite{jia2023} & 1 & $p_{1,15}$ \\
        16 & \textcite{wang2023b} & 1 & $p_{1,16}$ \\
        17 & \textcite{owers2022} & 1 & $p_{1,17}$ \\
        18 & \textcite{peng2022} & 1 & $p_{1,18}$ \\
        19 & \textcite{suryono2022} & 1 & $p_{1,19}$ \\
        $h = 20$ & \textcite{pareeth2019} & 1 & $p_{1,20}$ \\
        & & $m = \sum_j m_j = 86$ \\
        \bottomrule
    \end{tabular}}
\end{table}

\subsection{Features}

Using the selected papers and previous systematic reviews, a list of potential study features was created (Table \ref{tab:features}). The features provide information about the complexity of the classification tasks (e.g., the number of output classes) and the proportion of the majority class, indicating potential class imbalance issues that can affect the performance of classification models. In all papers, the labeled data were split randomly into training and test sets, so the proportion of the majority class does not systematically differ between sets. Remote sensing-specific information was also gathered, including the type of device, the number of spectral bands, and pixel resolution, to assess how data collection impacts performance. The latter two were categorized due to high levels of non-reporting. The number of citations was gathered using the Local Citation Network web app, which collects article metadata from OpenAlex---a bibliographic catalog of scientific papers \parencite{priem2022}.

\begin{table}[t!]
    \caption{Study-level features.}
    \label{tab:features}
    \centering
        \resizebox{\textwidth}{!}{

    \begin{tabular}{l p{0.4\linewidth} p{0.4\linewidth}}
        \toprule
        Feature $x_j$ & Description & Values \\
        \midrule
        \multicolumn{3}{l}{Categorical} \\
        \hspace{3mm}device & Remote sensing device & passive, active, combined, unknown \\
        \hspace{3mm}model & Type of ML model & neural network, tree-based, other \\
        \hspace{3mm}nbands & Number of spectral bands & [1,7], (7,14], unknown \\
        \hspace{3mm}resolution & Pixel resolution (m) & (0,10], (10,30], unknown \\
        \hspace{3mm}satellite & Type of satellite & Sentinel, Landsat, other, unknown \\
        \hspace{3mm}sdg & UN sustainable development goal & sustainable cities (SDG 11), life on land (SDG 15), zero hunger (SDG 2) \\
        \hspace{3mm}unit & Unit of classification & object, pixel, unknown \\
        \multicolumn{3}{l}{Numeric} \\
        \hspace{3mm}maxprev & Prevalence of majority class & 0.142--0.995 \\
        \hspace{3mm}ncitations & Number of citations & 0--68 \\
        \hspace{3mm}nclasses & Number of classes & 2--13 \\
        \hspace{3mm}year & Year of publication & 2018--2023 \\
        \multicolumn{3}{l}{Binary} \\
        \hspace{3mm}auxiliary & Indicator for the use of auxiliary data in addition to remote sensing data & 0,1 \\
        \hspace{3mm}confmatrix & Indicator for reporting a confusion matrix & 0,1 \\
        \hspace{3mm}indices & Indicator for the use of remote sensing indices & 0,1 \\
        \bottomrule
    \end{tabular}}
\end{table}

\subsection{Data exploration}
\label{sec:explore}

\begin{figure}[t!]
    \centering
    \includegraphics[width=\linewidth]{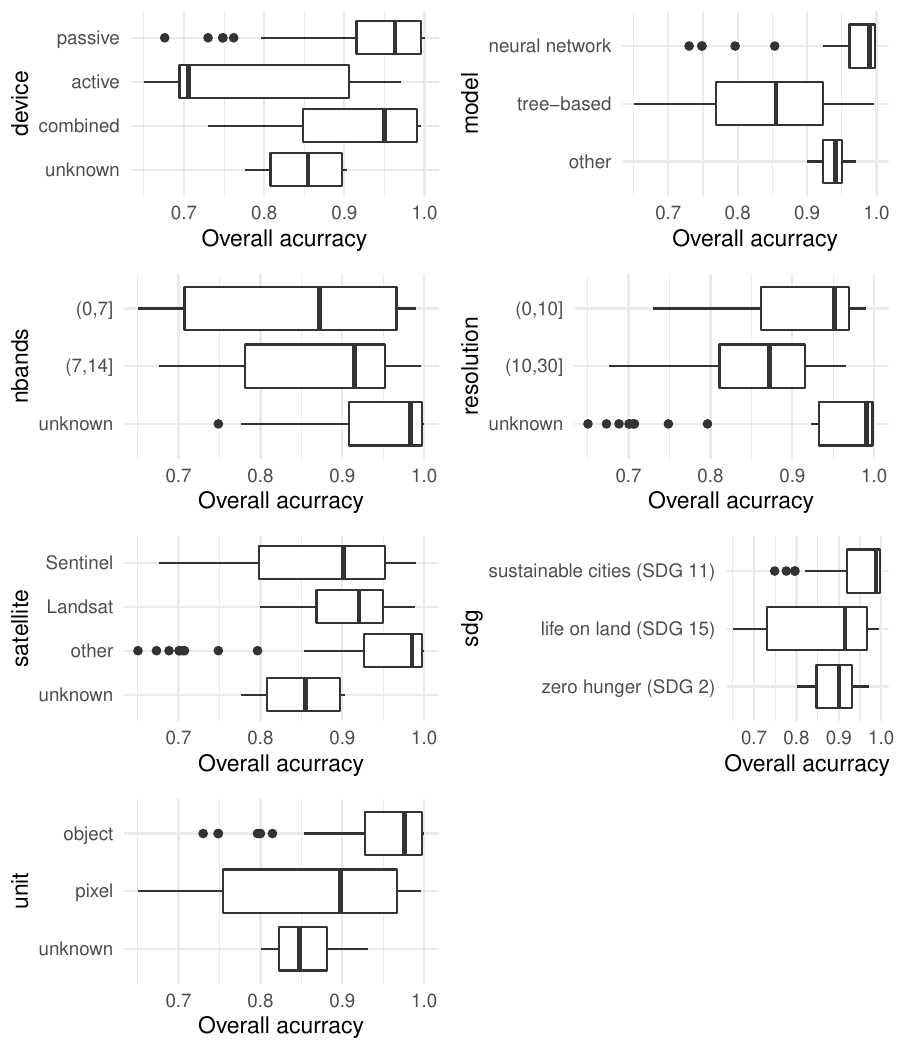}
    \caption{Distribution of overall accuracy by categorical feature.}
    \label{fig:OA_categorical}
\end{figure}

\begin{figure}[t!]
    \centering
    \includegraphics[width=0.7\linewidth]{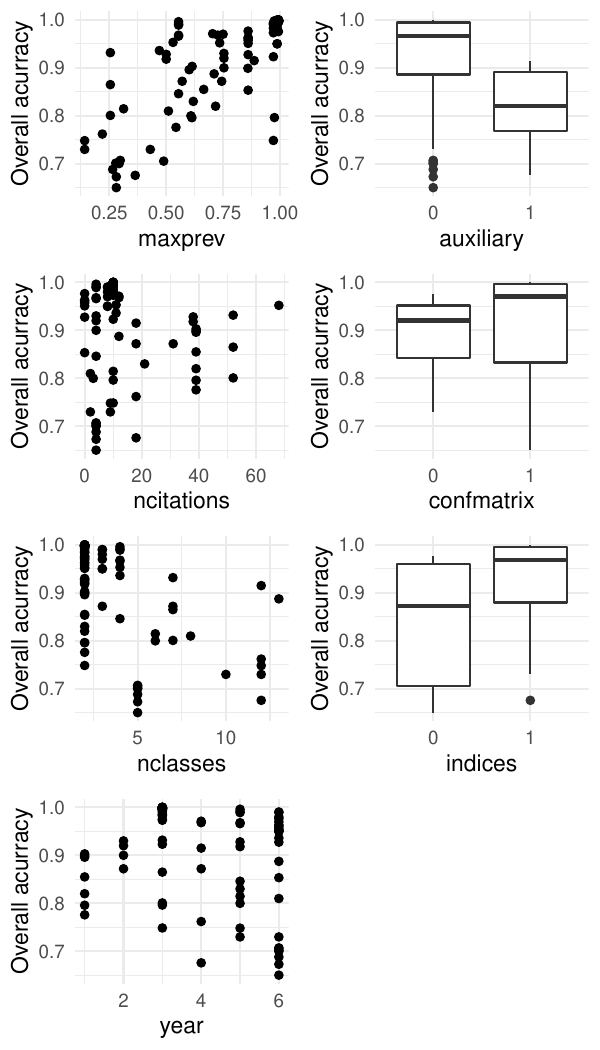}
    \caption{Distribution of overall accuracy by numeric feature (left column) and binary feature (right column).}
    \label{fig:OA_numeric_binary}
\end{figure}

For illustration purposes, the distribution of observed overall accuracy across trials is shown by categorical feature in Figure \ref{fig:OA_categorical} and by numeric feature and binary feature in Figure \ref{fig:OA_numeric_binary}. This data exploration is only observational, and no conclusions are drawn from it. Conclusions will be drawn from the statistical analysis described in Section \ref{sec:statanalysis}.

Each feature is described in Table \ref{tab:features}. The following specific observations considering the categorical feature in Figure \ref{fig:OA_categorical} can be made: Passive sensing tends to yield higher overall accuracy values than active or combined systems (device). Tree-based models and neural networks both achieve high accuracy ranges, while tree-based methods exhibit more variation (model). Regarding the number of bands (nbands), it seems that spectral richness improves model performance. However, the highest overall accuracies are found for the group where the number of bands is not reported. The different groups for the resolution do not show a clear pattern (resolution). Landsat and Sentinel satellites tend to perform similarly, with fewer variations for Landsat (satellite). However, other satellites outperform the two with some outliers. Studies in which the type of satellite remained unknown also performed well, with slight variations. The SDG 15 (Life on Land) exhibits a large variation (sdg), whereas studies regarding SDG 11 (Sustainable Cities) tend to have higher overall accuracies. SDG 2 (Zero Hunger) shows the slightest variation in overall accuracy. Concerning the unit, object-based classifications tend to yield higher overall accuracies than pixel-based ones (unit).

\noindent The following specific observations considering the numeric and binary features in Figure \ref{fig:OA_numeric_binary} can be made: a clear positive trend is observed between class imbalance, measured by maximum class prevalence (maxprev), and overall accuracy. This indicates that models trained on imbalanced datasets may achieve high accuracy by favoring dominant classes. The number of citations (ncitations) shows a modest correlation with higher accuracy, possibly reflecting methodological quality or publication bias. This feature is not a causal factor but a control variable, as a measure of a paper's impact or age. Conversely, the number of classes (nclasses) has an inverse relationship with accuracy, emphasizing the increased difficulty of more complex classification tasks. Over time, accuracy may have slightly improved, as indicated by the year variable. The use of auxiliary data significantly enhances performance, with models that incorporate external datasets outperforming those that do not. Similarly, the inclusion of spectral indices is associated with higher accuracy. Whether a study reports a confusion matrix (confmatrix) does not appear to influence reported accuracy, indicating that transparency in reporting does not necessarily correlate with performance outcomes.

To summarize, based on this visual inspection, the most effective setup may comprise passive sensors, neural networks, more spectral bands, Landsat imagery, object-based analysis, tasks related to SDG 11, models trained on imbalanced data, incorporation of auxiliary data and indices, and a non-complex classification task. This will be tested using the statistical analysis described in the next section.

\clearpage
\subsection{Statistical analysis}\label{sec:statanalysis}

\subsubsection*{Transformation}

Before analyses, the \textcite{freeman1950} double arcsine transformation ($\DA$) was applied to the observed overall accuracy, $p_{ij}$, resulting in the observed individual effect size, $\tilde{\theta}_{ij}$, and its sampling variance $v_{ij}$:

\begin{align}
    p_{ij} &= \frac{k_{ij}}{n_{ij}}, \\
    \tilde{\theta}_{ij} &= \DA(p_{ij}) = \frac{\sin^{-1} \sqrt{\frac{k_{ij}}{n_{ij} + 1}} + \sin^{-1} \sqrt{\frac{k_{ij} + 1}{n_{ij} + 1}}}{2}, \\
    v_{ij} &= 1 / (4n_{ij} + 2).
\end{align}

\noindent where $k_{ij}$ is the number of correctly classified instances and $n_{ij}$ the number of instances in trial $i$ of study $j$. After some controversy, this transformation is preferred for the meta-analysis of proportions \parencite{doi2021}. No transformation, or other transformations, like log, logit, and arcsine square root, yielded higher deviations from normality than DA according to the Shapiro-Wilk test.

\subsubsection*{Three-level meta-analysis}

A multilevel linear model was fitted to estimate the population effect size and the heterogeneity between and within studies. This model takes into account that individual effect sizes are not independent within a study.

At individual level (1), $\tilde{\theta}_{ij}$ is the observed individual effect size from trial $i$ in study $j$. It is modeled as the (unobserved) true individual effect size $\theta_{ij}$ and an error component $\epsilon_{ij}$. The error component is assumed to be normally distributed with mean 0 and sampling variance $v_{ij}$:

\begin{align}
    \tilde{\theta}_{ij} &= \theta_{ij} + \epsilon_{ij}, \nonumber \\
    \epsilon_{ij} &\sim N(0, v_{ij}).
\end{align}

At study level (2), the true individual effect size $\theta_{ij}$ from trial $i$ in study $j$ is modeled as the study effect size $\kappa_j$ in study $j$ and within-study heterogeneity $\zeta_{ij}$. This heterogeneity is assumed to be normally distributed with mean 0 and within-study variance $\sigma^2_\zeta$:

\begin{align}
    \theta_{ij} &= \kappa_j + \zeta_{ij}, \nonumber \\
    \zeta_{ij} &\sim N(0, \sigma^2_\zeta).
\end{align}

\noindent Note that the within-study variance $\sigma^2_\zeta$ is assumed constant across studies, which is a common assumption. At population level (3), the study effect size $\kappa_j$ in study $j$ is modeled as the population effect size, $\mu$ and between-study heterogeneity $\xi_j$. This heterogeneity is assumed to be normally distributed with mean 0 and between-study variance $\sigma^2_\xi$:

\begin{align}
    \kappa_j &= \mu + \xi_j, \nonumber \\
    \xi_j &\sim N(0, \sigma^2_\xi).
\end{align}

Combined, the observed individual effect size is decomposed into the population effect size, the between-study heterogeneity, the within-study heterogeneity, and an error component: $\tilde{\theta}_{ij} = \mu + \xi_j + \zeta_{ij} + \epsilon_{ij}$. Its total variance is $\mathrm{Var}(\tilde{\theta}_{ij}) = \sigma^2_\xi + \sigma^2_\zeta + v_{ij}$.

The population effect size $\mu$ is a weighted sum of the $m = \sum_j m_j$ observed individual effect sizes $\tilde{\theta}_r$ \parencite{konstantopoulos2011,cheung2014}:

\begin{align}
    \mu &= \frac{\sum_{r=1}^m (\sum_{c=1}^m w_{rc}) \tilde{\theta}_r}{\sum_{rc} w_{rc}}, \\
    \mathrm{Var}(\mu) &= \frac{1}{\sum_{rc} w_{rc}}.
\end{align}

\noindent The weight $w_{rc}$ is not simply the inverse variance $1 / (\sigma^2_\xi + \sigma^2_\zeta + v_{ij})$ but takes into account the covariance between individual effect sizes within a study. It is the entry in the $r$-th row and $c$-th column of an $m \times m$ weight matrix $\bm{W}$, which is the inverse of the marginal variance-covariance matrix $\bm{M}$. This matrix $\bm{M}$ contains $\sigma^2_\xi + \sigma^2_\zeta + v_{ij}$ on the diagonal and $\sigma^2_\xi$ on the off-diagonal within each study:

\begin{align}
    \bm{W} &= \bm{M}^{-1}, \\
    \bm{M} &= \oplus_{j = 1}^{h} \left( \bm{J}_{m_j} \sigma^2_\xi + \bm{I}_{m_j} \sigma^2_\zeta + \bm{I}_{m_j} v_{ij} \right),
\end{align}

\noindent where $\oplus$ is the matrix direct sum, $\bm{J}_{m_j}$ an all-ones matrix of order $m_j$, and $\bm{I}_{m_j}$ an identity matrix of order $m_j$. Thus, individual effect sizes in different studies are assumed independent ($\mathrm{Cov}(\tilde{\theta}_{ij}, \tilde{\theta}_{kl}) = 0$), and individual effect sizes within a study share the same covariance ($\mathrm{Cov}(\tilde{\theta}_{ij}, \tilde{\theta}_{kj}) = \sigma^2_\xi$) \parencite{konstantopoulos2011,cheung2014}. Parameters are estimated using restricted maximum likelihood (REML). Other options exist, but REML is suitable if the target variable is continuous.

To assess the relative contributions of the three variance components to the total variance, the total variance is simplified to $\mathrm{Var}(\tilde{\theta}) = \sigma^2_\xi + \sigma^2_\zeta + \sigma^2_\epsilon$, where $\sigma^2_\epsilon$ is a pooled estimate of the $m$ sampling variances $v_{ij}$ \parencite{higgins2002}:

\begin{equation}
    \sigma^2_\epsilon = \frac{(m - 1) \sum_{ij} 1 / v_{ij}}{(\sum_{ij} 1 / v_{ij})^2 - \sum_{ij} 1 / v_{ij}^2}.
\end{equation}

\subsubsection*{Three-level meta-regression}

The variance between and within studies might be explained by the features (aka covariates, predictors, independent variables, moderators). To test this, the multilevel model is extended in the following way. Let $\bm{x}_j$ be a vector of $d$ features about study $j$. Sub-study-level features $\bm{x}_{ij}$ could also be included, but are ignored here since we do not have any. The model becomes:

\begin{equation}
    \label{eq:model}
    \tilde{\theta}_{ij} = \mu + \bm{x}_j^\intercal \bm{\beta}_j + \xi_j + \zeta_{ij} + \epsilon_{ij},
\end{equation}

\noindent where $\bm{\beta}_j$ is a vector of $d$ regression coefficients. The heterogeneity $\xi_j + \zeta_{ij}$ is now the variability across the true individual effect sizes $\theta_{ij}$ that is not explained by the features. Relevant features were selected using a forward selection procedure and Akaike Information Criterion (AIC) to find the right balance between model fit and parsimony \parencite{Venables_Ripley_2002}. Any decrease in the information criterion was considered an improvement.

Adding features to the model will reduce the variances at population level ($\xi$) and study level ($\zeta$). The reduction is quantified by $R^2$:

\begin{align}
    R^2_\xi &= 1 - \frac{\sigma^2_\xi(X)}{\sigma^2_\xi(0)}, \\
    R^2_\zeta &= 1 - \frac{\sigma^2_\zeta(X)}{\sigma^2_\zeta(0)},
\end{align}

\noindent where $\sigma^2(X)$ is the variance according to the model with features and $\sigma^2(0)$ the variance according to the model without features (the null model).

\subsubsection*{Backtransformation}

Modeling is done on the DA-transformed proportions, but the results are backtransformed for ease of interpretation. The backtransformation of the \textcite{freeman1950} double arcsine transformation was initially proposed by \textcite{miller1978}. To avoid numerical instability, \textcite{barendregt2013} replaced the harmonic mean of sample sizes ($\hat{n} = m / \sum_{ij} n_{ij}^{-1}$) by the inverse population variance ($\hat{n} = 1 / \mathrm{Var}(\mu) = \sum_{rc} w_{rc}$) as an estimate of the sample size $n$:

\begin{align}
    \DA^{-1}(\gamma) &= \frac{1 - \sgn \left( \cos(\gamma) \right) \sqrt{1 - \left[\sin(\gamma) + \frac{\sin(\gamma) - \frac{1}{\sin(\gamma)}}{\hat{n}} \right]^2}}{2}, \\
    \overline{p} &= \DA^{-1}(2 \mu), \\
    \mathrm{LCB}(\overline{p}) &=
    \begin{cases}
        0 & \text{if } \overline{p} \hat{n} < 2  \\
        \DA^{-1} \Bigl( 2 \bigl[ \mu - t_{1 - \alpha / 2, h - 1} \sqrt{\mathrm{Var}(\mu)} \bigr] \Bigr) & \text{otherwise}
    \end{cases}, \\
    \mathrm{UCB}(\overline{p}) &=
    \begin{cases}
        1 & \text{if } (1 - \overline{p}) \hat{n} < 2 \\
        \DA^{-1} \Bigl( 2 \bigl[ \mu + t_{1 - \alpha / 2, h - 1} \sqrt{\mathrm{Var}(\mu)} \bigr] \Bigr) & \text{otherwise}
    \end{cases},
\end{align}

\noindent where $\gamma$ is any double arcsine transformed proportion, $\sgn$ the sign function, $\overline{p}$ the backtransformed population effect size, LCB the lower confidence bound, UCB the upper confidence bound, and $t_{1 - \alpha, \nu}$ the critical value of Student's t distribution with significance level $\alpha$ and $\nu$ degrees of freedom. Using the inverse population variance as the effective sample size provides greater stability than the harmonic mean because it weights studies by the precision of their estimates rather than being disproportionately influenced by small sample sizes. For the model with study-level features, $\overline{p}$ is replaced by $\overline{p}_j$, $\mu$ by $\mu + \bm{x}_j^\intercal \bm{\beta}_j$, and $\hat{n}$ by $\hat{n}_j$.

\subsubsection*{Benchmark}

Suppose one simply guesses each class $r$ with a probability equal to its prevalence $n_{r.}/n$ in the training set (see Section \ref{sec:ML}). The benchmark overall accuracy then equals:

\begin{equation}\label{eq:benchmark}
   p_0 = \sum_{r = 1}^q \left(\frac{n_{r.}}{n} \right)^2.
\end{equation}

\noindent For instance, if two classes are in balance, then $p_0 = 0.5$. But if ten classes are in balance (uniformly distributed), then $p_0 = 0.1$. And if two classes are imbalanced with odds 9 to 1, then $p_0$~=~0.82. In general, the fewer classes and the more class imbalance, the higher the benchmark and the easier it is to get high overall accuracy scores. Thus, when analyzing overall accuracy across studies, it is important to incorporate the number of classes and the amount of class imbalance as features. An alternative would be to normalize overall accuracy with its benchmark, but this is not common practice. Note that for a fair test result, one needs to ensure, for example by weighting, that the class distributions in the training and test set are unbiased estimates of the class distribution in the population.

\subsubsection*{Implementation}

For the statistical analyses, the following R functions were used: \texttt{escalc}, \texttt{vcalc}, \texttt{rma.mv}, \texttt{rma}, \texttt{forest} and \texttt{regplot} from the package \texttt{metafor} \parencite{viechtbauer2010}, and \texttt{var.comp} from the package \texttt{dmetar} \parencite{harrer2019}.

\section{Results}\label{sec:results}

\subsection{Three-level meta-analysis}

Table \ref{tab:summary} summarizes the results. According to Cochran’s Q test on heterogeneity, the variability in the observed effect sizes is larger than one would expect based on the sampling variances alone. This suggests that the true individual effect sizes are heterogeneous. Without including features, the between-study variance $\sigma^2_\xi$ accounts for 64\% of the total variance $\mathrm{Var}(\tilde{\theta}_{ij})$, while the remainder is attributed to the within-study variance $\sigma^2_\zeta$, making the contribution of the pooled sampling variance $\tilde{v}$ negligible. The population effect size on the double arcsine scale, $\mu$, and its variance are also given in Table \ref{tab:summary}, but it is easier to interpret after backtransformation to $\overline{p}$: the population estimate of the quality metric overall accuracy is estimated at 0.89 with a 95\% confidence interval of 0.85--0.93.

\begin{figure}[t!]
    \centering
    \includegraphics[width=\linewidth]{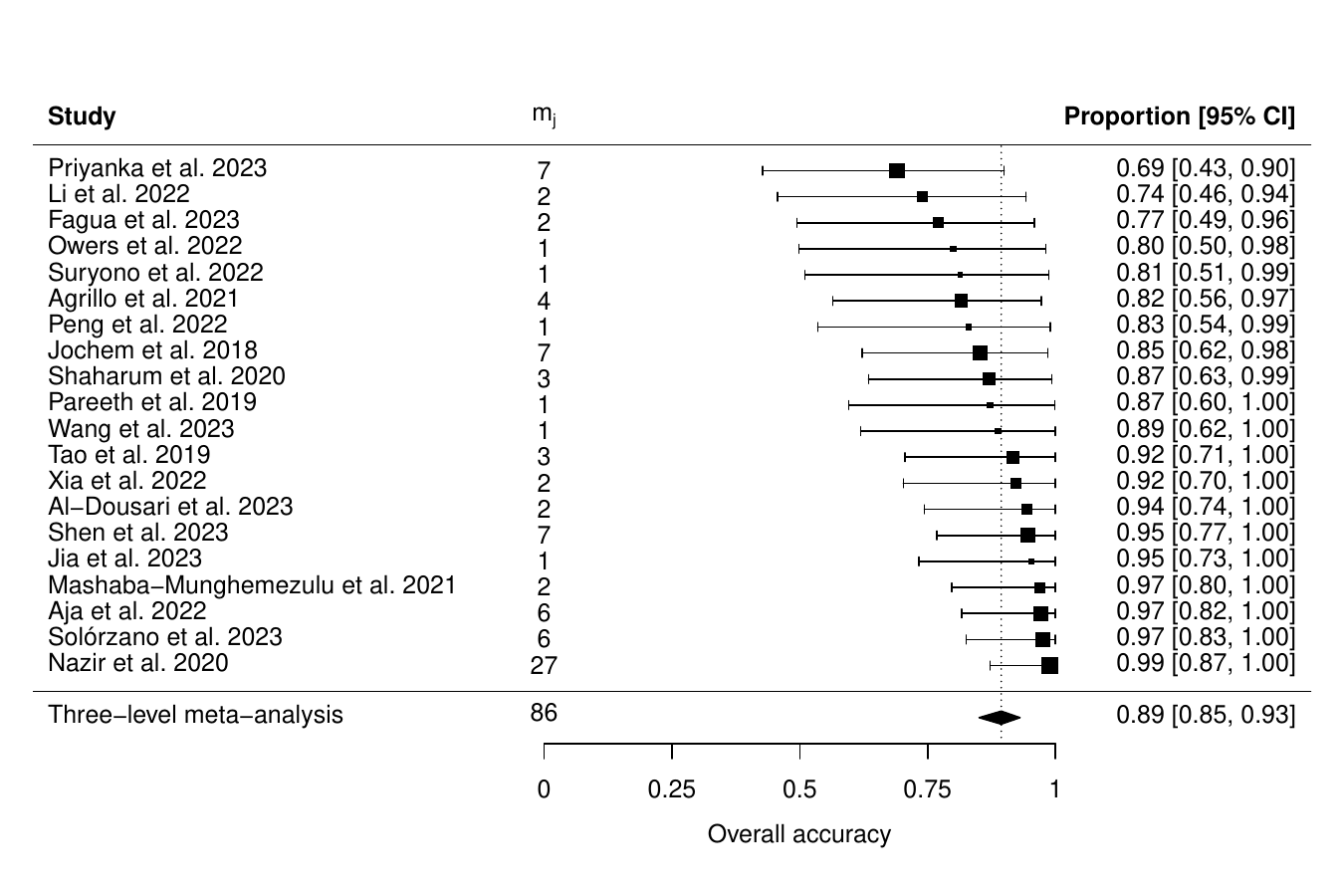}
    \caption{Forest plot of meta-analysis.}
    \label{fig:forest_analysis}
\end{figure}

Figure \ref{fig:forest_analysis} shows a forest plot. Each row represents one of the $h = 20$ studies with its $m_j$ number of trials. Next, $\overline{p}$ and the $h = 20$ study-level estimates $\kappa_j$ of the quality metric, overall accuracy, along with their confidence intervals, are shown. The size of the square, which indicates the value of $\kappa_j$, is determined by the size of $j$, i.e., the more trials per study, the larger the square. The population estimate of the quality metric overall accuracy is shown at the bottom line. The dashed line indicates the population estimate, which is also covered by the confidence intervals of all studies.

\begin{table}[t!]
    \caption{Summary of fitted models. The $I^2$ and $R^2$ statistics are reported as a fraction.}
    \label{tab:summary}
    \centering
    \begin{tabular}{lll}
        \toprule
         & Meta-analysis & Meta-regression \\
        \midrule
        $Q$ & $1.2 \times 10^7$ ($p < .0001$) & $ 1.1 \times 10^7$ ($p < .0001$) \\
        df & 85 & 84 \\
        $\sigma^2_\xi$ ($h = 20$) & $0.0173$ ($I^2_\xi = 0.64$) & $0.0068$ ($I^2_\xi = 0.43$) \\
        $\sigma^2_\zeta$ ($m = 86$) & $0.0099$ ($I^2_\zeta = 0.36$) & $0.0091$ ($I^2_\zeta = 0.57$) \\
        $\sigma^2_\epsilon$ & $10^{-8}$ ($I^2_\epsilon = 10^{-6}$) & $10^{-8}$ ($I^2_\epsilon = 10^{-6}$) \\
        $R^2_\xi$ & & $0.61$ \\
        $R^2_\zeta$ & & $0.08$ \\
        $\mu$ (SE) & $1.2384$ ($0.0331$) & $0.9796$ ($0.0557$) \\
        $\beta(\mathrm{maxprev})$ (SE) & & $0.4335$ ($0.0833$) \\
        \bottomrule
    \end{tabular}
\end{table}

\subsection{Three-level meta-regression}

\begin{figure}[htbp]
    \centering
    \includegraphics[width=.95\linewidth]{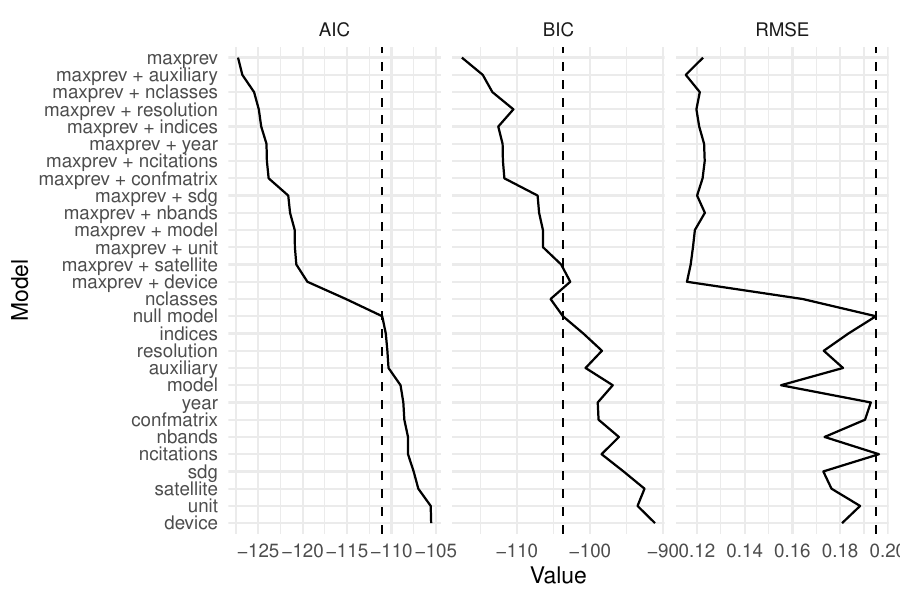}
    \caption{Forward feature selection. Models are sorted by AIC. Dashed lines show values of the null model (intercept only). More complex models were not fitted because adding a feature to a model with only the maximum prevalence (maxprev) did not lower the AIC. X-axis shows the AIC, BIC, or RMSE values. Lower AIC and BIC values are related to better balanced models regarding fit and complexity. The RMSE informs on predictive performance, but does not take into account the complexity.}
    \label{fig:aic}
\end{figure}

Figure \ref{fig:aic} shows the result of the applied forward model selection. The y-scale of the plot shows the features and feature combinations of the forward model selection. The x-scale shows the values of the different quality metrics. Each panel shows a different quality metric. For evaluation, three widely used model evaluation metrics were chosen: Akaike Information Criterion (AIC), Bayesian Information Criterion (BIC), and Root Mean Square Error (RMSE). AIC balances the trade-off between goodness of fit and model complexity. BIC is similar to AIC but applies a stricter penalty for model complexity. The RMSE is the average magnitude of prediction errors, i.e., the average distance between predicted values and actual values. By using these three different metrics, we examine the models from different angles, aiming to find those that are not only accurate but also parsimonious and generalizable.

For all quality metrics, the prevalence of the majority class (maxprev) had a significant effect on the reported overall accuracy. For the AIC and BIC, adding any additional feature did not lower their values (lower AIC or BIC values indicate a better model). According to the RMSE, either the indicator for the use of auxiliary data or the remote sensing device (e.g., passive, active) could be added. Still, this metric focuses on model fit and does not account for model complexity. We also observe that for AIC and BIC, several models that use only a single feature perform worse compared to the null model (those features with values to the right of the dashed line, as the dashed line represents the value of the null model). Thus, these features only add noise and do not pick up any signal in the data.

\noindent In addition, a permutation test was conducted to evaluate the robustness of the applied feature selection methods. The most important finding is that the permutation feature importance \parencite{molnar2025} confirmed that the prevalence of the majority class (maxprev) is the most important feature (see Appendix \ref{app:permutation}).

\begin{figure}[htbp]
    \centering
    \includegraphics[width=.95\linewidth]{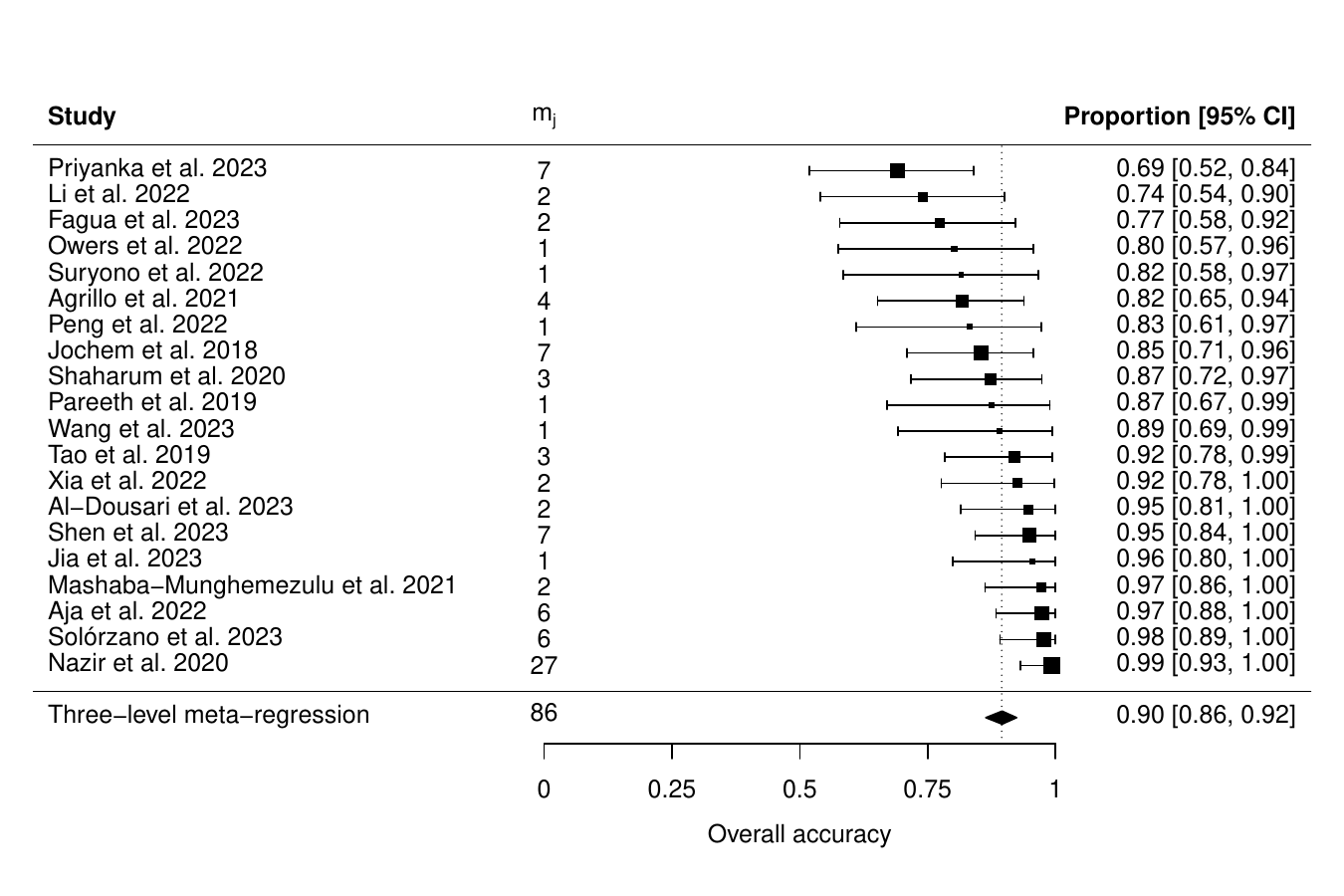}
    \caption{Forest plot of meta-regression, including the prevalence of the majority class (maxprev) as only significant feature.}
    \label{fig:forest_regression}
\end{figure}

Including maxprev as a feature in the model does not change the result of Cochran's Q test for heterogeneity (Table \ref{tab:summary}). This heterogeneity is now partly explained by the feature maxprev. Adding maxprev to the model dropped the proportion of the total variance explained by the between-study variance from 64\% to 43\% ($I^2_\xi$), a relative reduction of 61\% ($R^2_\xi$). Since maxprev is a study-level feature, its effect on the within-study variance is much smaller ($R^2_\zeta$~=~8\%). Adding maxprev to the model ($\beta(\mathrm{maxprev})$ in Table \ref{tab:summary}) also narrows down the confidence intervals of both the population estimate (for the average maxprev) and the $h = 20$ study-level estimates $\kappa_j$. This is illustrated in Figure \ref{fig:forest_regression}, which can be compared with Figure \ref{fig:forest_analysis}. The population estimate of the quality metric overall accuracy is now estimated at 0.90 with a 95\% confidence interval of 0.86--0.92. Still, most confidence intervals of the studies cover the population estimate. However, there is now one study with a significantly smaller estimate \parencite{priyanka2023} and one study with a significantly larger estimate \parencite{nazir2020}. We conducted a leave-one-study-out analysis by excluding \textcite{nazir2020} with 27 trials. The model selection shows that the best model according to AIC includes maxprev and resolution, whereas the best model according to BIC still includes only maxprev. Accordingly, the main conclusion from the paper would not change without \textcite{nazir2020}.

\begin{figure}[ht!]
    \centering
    \includegraphics[width=\linewidth]{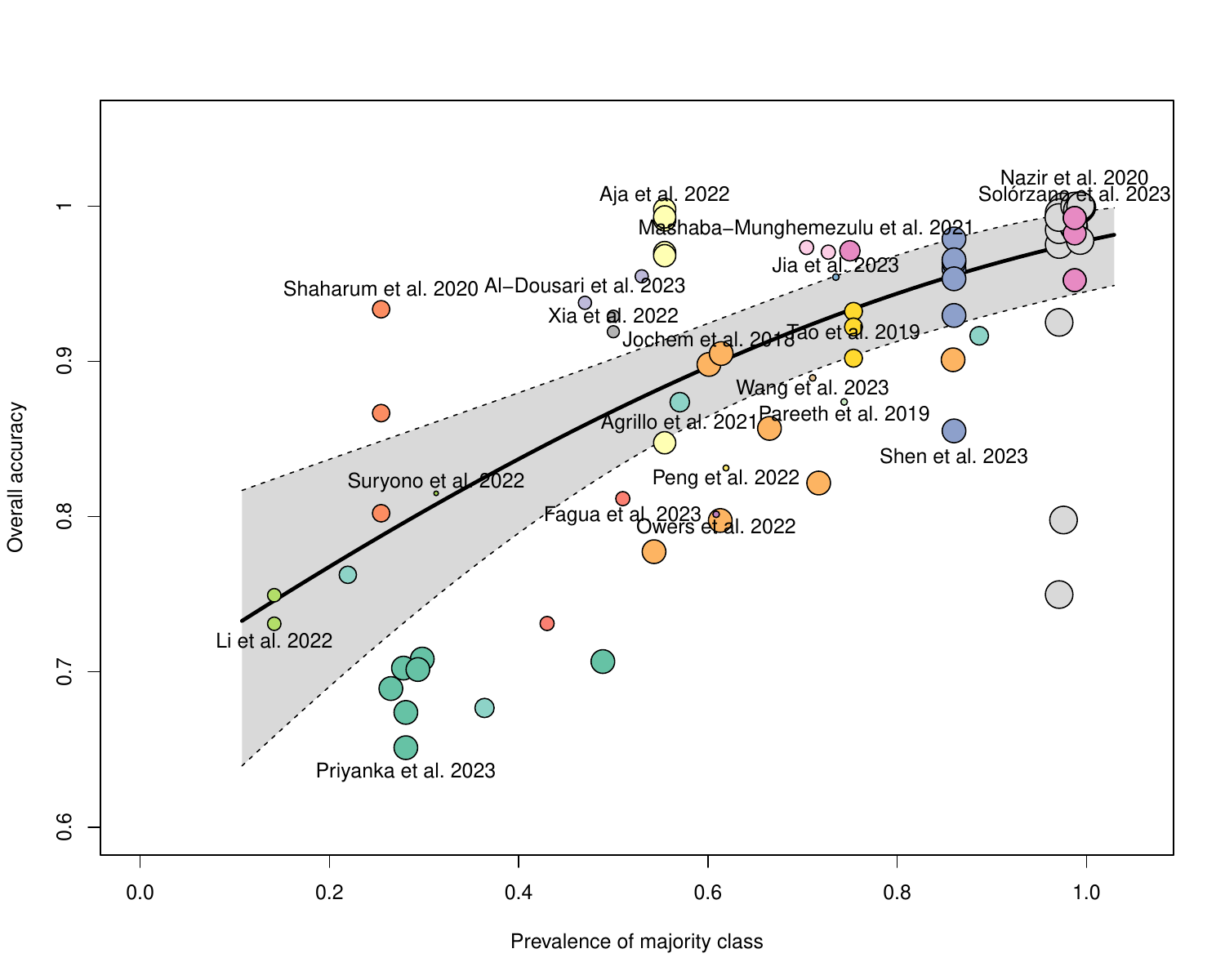}
    \caption{Overall accuracy against prevalence of the majority class for $m = 86$ trials within $h = 20$ studies. Points are colored by study, and their size is a function of the model weights. Regression line and 95\% confidence interval result from the DA-backtransformed three-level meta-regression model. Note that the y-axis does not start at zero.}
    \label{fig:regplot}
\end{figure}

\noindent The overall accuracy increases with the prevalence of the majority class. The regression coefficient on the DA scale is positive (Table \ref{tab:summary}). In Figure \ref{fig:regplot} this is illustrated on the backtransformed scale. The y-scale shows the backtransformed overall accuracy and the x-scale the prevalence of the majority class. The $h = 20$ studies with their $m = 86$ trials are plotted. The solid black line represents the regression line with its 95\% confidence intervals (dashed lines). Thus, the more dominant one class, the higher the overall accuracy. This finding was expected and is theoretically explained in Section \ref{sec:statanalysis}, Equation \ref{eq:benchmark}. 

\section{Discussion}\label{sec:discussion}

This meta-analysis evaluated the performance of ML models for SDG monitoring using remote sensing data. Specifically, the study estimated the average performance, determined the level of heterogeneity between and within studies, and assessed which study features are related to model performance.

The results show that the overall accuracy is consistently high. The population overall accuracy was estimated at $\overline{p} = 0.89$ $[0.85, 0.93]$ without features and more precisely at $0.90$ $[0.86, 0.92]$ with the prevalence of the majority class (maxprev) as only significant feature included. The results also demonstrate considerable heterogeneity, 64\% of which was between studies, 61 \% of which could be explained by the maxprev feature. None of the other thirteen study features had a significant additional effect. This contrasts \textcite{khatami2016} and \textcite{hanade2022} who found that using auxiliary data did improve model performance. It also contrasts \textcite{khatami2016} who found some effects of the type of ML model. Notably, no study was found that explicitly corrected for class imbalance (proportion of the majority class) when assessing the difference in performance between groups. The effects of the other features were overshadowed by the strong effect of the prevalence of the majority class.This might explain the differences between our study and the studies by \textcite{khatami2016} and \textcite{hanade2022}.

\subsubsection*{Limitations and future research}

Overall accuracy is the most commonly reported metric, which was also found by \textcite{khatami2016,jafarzadeh2022}. However, overall accuracy has several limitations \parencite{foody2020,stehman2019}. Most importantly, it is sensitive to class imbalance, which compromises between-study comparisons. This explains why the proportion of the majority class was a significant predictor. Reporting arithmetic mean recall (balanced accuracy) or geometric mean recall would be an improvement. Other metrics like Matthew's Correlation Coefficient (MCC) or Cohen's Kappa should be considered as well. Reporting several metrics normalized with a benchmark \parencite{burger2020} would be even better, or reporting a full confusion matrix from which all these metrics can be derived. Unfortunately, this is far from common practice yet. Multiple performance metrics can be modeled using network meta-analysis \parencite{harrer2021}, which would be an interesting topic for future research. The focus on overall accuracy may provide policymakers a too positive picture of the progress regarding the SDGs. Future work could also try to incorporate sub-study-level features, as they were not collected in this study. Such features could be, for example, specific ML algorithms or hyperparameter configurations.

Of the 200 studies randomly sampled, only 57 papers were agreed upon by all three reviewers, while each reviewer thought between 77 and 81 studies could have been included. This highlights the subjectivity of the selection process (probably due to vague abstracts and interdisciplinary scope) and the importance of having multiple reviewers. Efficiency could be gained from using the open-source tool ASReview \parencite{schoot2021} and large language models for feature extraction \parencite{mahuli2023}, but these cannot solve inconsistent or unclear reporting on methods in this field. For example, it was unclear in some of the selected studies whether the performance was reported on the training set, the validation set, the test set, or the entire labeled dataset. Furthermore, Google Scholar was excluded for good reasons (see Section \ref{sec:selection}), but the impact of this decision has not been evaluated.

This analysis included a total of 86 trials from 20 studies. The low yield (10\%) from the initial sample of 200 is caused by carefully filtering out irrelevant papers and illustrates the difficulty and labor-intensity of compiling an appropriate dataset. While several simulation studies suggest that a three-level meta-analysis can yield accurate results with as few as 20 to 40 studies \parencite{hedges2010,polanin2014}, this analysis is at the lower bound. Reviewing more papers will increase the statistical power and open up possibilities for testing interactions between study features, dose-response meta-analysis \parencite{viechtbauer2024}, and robust variance estimation for correlated sampling variances \parencite{harrer2021}.

This study only examined published results, which introduces publication bias---a well-documented effect where studies with positive results are more likely to be published, while negative or neutral findings remain unpublished \parencite{borenstein2009,bozada2021,hansen2022,harrer2021}. This bias can lead to an overestimation of the population overall accuracy. Our decision to include only papers that report overall accuracy excludes studies that responsibly chose not to report overall accuracy due to its sensitivity to class imbalance. This could also introduce publication bias, although this bias more likely leads to an underestimation of the population overall accuracy. Another source of potential publication bias, that cannot be ultimately checked, is that studies likely report only the performance of the best model, ignoring the performance of other tested models. The paper does not quantify potential publication bias. The use of funnel plots in the context of meta-analyses of proportions is debated. We provide a funnel plot in Appendix \ref{app:funnel}. Conventional funnel plot methodology is inaccurate for meta-analyses of proportion studies \parencite{hunter2014}. The most common tests for assessing potential publication bias are Egger’s test and Begg’s test. However, these methods are also problematic in the context of hierarchical data. They assume that each study contributes a single, independent effect size estimate \parencite{sutton2009, rodgers2020}, and thus would also lead to misleading conclusions. Moreover, funnel plot asymmetry can be caused by reasons other than publication bias, such as between-study heterogeneity \parencite{fernandez2020}.

\subsubsection*{Conclusion}

The monitoring and implementation of the United Nations’ Sustainable Development Goals (SDGs) is of high practical importance to tackle current societal challenges. For this purpose, ML is a widely adopted and used tool. Overall accuracy of ML models applied to remote sensing data can be considered good. However, overall accuracy is a popular yet poor quality metric for between-study comparisons. Overall accuracy is sensitive to class imbalance, and studies differ in the degree of class imbalance. Thus, using overall accuracy as a single quality metric could lead to a too optimistic picture. Policymakers and international organizations must consider these findings and their policy implications in real-world SDG monitoring. 

For example, in projects that monitor deforestation rates (SDG 15) or urban expansion (SDG 11), relying solely on overall accuracy may mask the poor detection of minority classes, such as small patches of forest loss or informal settlements. This can lead to misguided policy decisions, ineffective allocation of resources, and missed opportunities for timely interventions. Therefore, if a project for SDG monitoring is funded, it must be contractually stipulated that comprehensive model performance metrics, such as confusion matrices and class-specific accuracies, are reported in addition to overall accuracy. Using more normalized metrics, such as F1-score, precision, recall, or Cohen’s Kappa, can provide a more balanced evaluation of model reliability.

These quality assurance measures are crucial to ensuring the transparency, comparability, and trustworthiness of SDG monitoring efforts. By adopting these standards, policymakers can make more informed decisions, tailor interventions more effectively, and ultimately enhance progress toward achieving the SDGs. Without such rigorous performance reporting requirements, there is a risk that suboptimal models could be used, undermining the credibility and impact of SDG monitoring initiatives.


        	\section*{Disclaimer}
	
	
	\noindent The views expressed in this report are those of the authors and do not necessarily correspond to the policies of Statistics Netherlands.
	
	\printbibliography

\newpage
\begin{appendices}

\section{Permutation feature importance}
\label{app:permutation}

Permutation feature importance was calculated according to \textcite[][Chapter 23]{molnar2025}. Feature importance is calculated as the proportional increase in RMSE when a feature is permuted, i.e. randomly shuffled. The stronger the effect of breaking the relationship between the feature and the target variable on the prediction error, the more important the feature.

To get an out-of-sample prediction error, the data is first randomly split into $K = 5$ folds. The model (Eq. \ref{eq:model}) with all $d = 14$ features (Table \ref{tab:features}) is trained on $K - 1 = 4$ folds (the training set) and used to predict the target variable for the remaining fold (the test set). Comparing the predicted with the observed individual effect sizes $\tilde{\theta}_{ij}$ yields the original RMSE. A feature is then permuted in a copy of the test set and the same trained model is used to predict the target variable, yielding the permuted RMSE. Permutation feature importance (PFI) is defined as the ratio between the permuted RMSE and the original RMSE. For each fold, the model is retrained and each feature is permuted $B = 200$ times (out of $(m / K)! = (86/5)! \approx 10^{14}$ possible permutations).

\begin{figure}[hb!]
    \centering
    \includegraphics[width=0.5\linewidth]{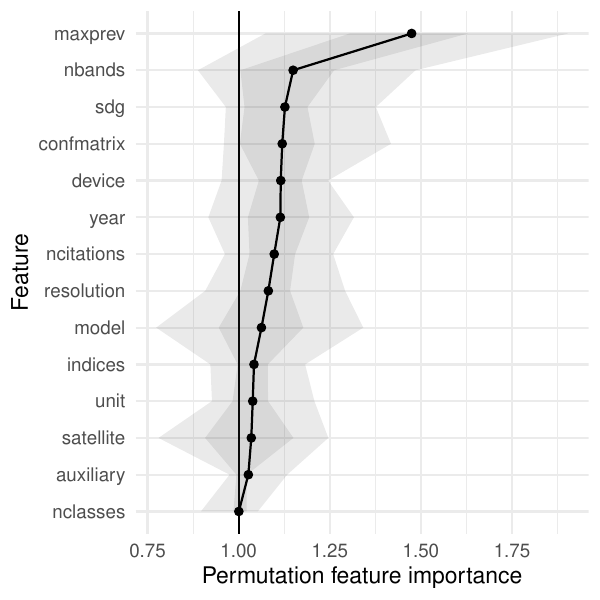}
    \caption{Permutation feature importance. Means (points), 25th and 75th percentiles (dark gray), and 2.5th and 97.5th percentiles (light gray) across 5 folds $\times$ 100 replicates. The vertical line indicates PFI = 1, i.e. no effect.}
    \label{fig:permutation}
\end{figure}

Figure \ref{fig:permutation} shows per feature the distribution of the PFI across $KB = 1000$ replicates. The features are ordered by mean PFI. The most important feature according to this method is the maximum prevalence of the majority class (maxprev). It is the only feature for which the 2.5th percentile of the PFI exceeds 1. This corroborates the feature selection methods using penalized likelihood (Fig. \ref{fig:aic}). Permuting maxprev increases the RMSE on average by a factor 1.47. The importance of the other features does not corroborate Figure \ref{fig:aic}. Note that Figure \ref{fig:aic} shows the effect of adding a single feature to the null model on the penalized likelihood, whereas Figure \ref{fig:permutation} shows the effect of permuting a feature on the prediction error of the full model. A disadvantage of the PFI method is that random shuffling of a feature can create unrealistic feature combinations and hence unrealistic predictions.

\clearpage
\section{Funnel plot}
\label{app:funnel}

\begin{figure}[htb!]
    \centering
    \includegraphics[width=0.6\linewidth]{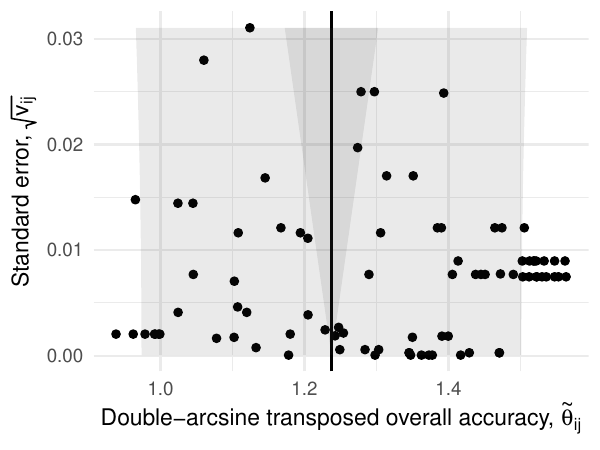}
    \caption{Funnel plot. Each point is an estimate of trial $i$ within study $j$, the vertical line is the population effect size ($\mu$), the shaded regions indicate 95\% confidence intervals $\mu \pm t_{0.975, h - 1} \sqrt{Var(\tilde{\theta}_{ij})}$, with only sampling variance ($Var(\tilde{\theta}_{ij}) = v_{ij}$, dark gray) and all variance components ($Var(\tilde{\theta}_{ij}) = \sigma^2_\xi + \sigma^2_\zeta + v_{ij}$, light gray).}
    \label{fig:funnel}
\end{figure}

As stated in the discussion, the use of funnel plots in the context of meta-analyses of proportions is a topic of debate. However, a funnel plot has been generated (see Figure \ref{fig:funnel}). One may speculate that there is a slight positive publication bias. According to \textcite{fernandez2020}, no definite conclusions can be drawn from this plot. We recommend future research to focus on developing publication-bias methodology in the context of meta-analyses of proportions and hierarchical data.

\end{appendices}

\end{document}